\documentclass[runningheads]{cl2emult}

\usepackage{makeidx}  % allows index generation
\usepackage{graphicx} % standard LaTeX graphics tool
\usepackage{subeqnar} % subnumbers individual equations
\usepackage{multicol} % used for the two-column index
\usepackage{cropmark} % cropmarks for pages without
\usepackage{eso}      % placeholder for figures
\makeindex            % used for the subject index
\def\ie{{\it i.e.,~}}
\def\eg{{\it e.g.,~}}
\def\etal{{\it et al.~}}
\def\cf{{\it cf.~}}
\def\etc{{\it etc.}}
\def\lsim{\hbox{ \rlap{\raise 0.425ex\hbox{$<$}}\lower 0.65ex\hbox{$\sim$} }}
\def\gsim{\hbox{ \rlap{\raise 0.425ex\hbox{$>$}}\lower 0.65ex\hbox{$\sim$} }}
\begin{document}
\title*{Exploration of Large Digital Sky Surveys}
\toctitle{Exploration of Large Digital Sky Surveys}
% allows explicit linebreak for the table of content
%
%
\titlerunning{Exploration of Sky Surveys}
% allows abbreviation of title, if the full title is too long
% to fit in the running head
%
\author{S.G. Djorgovski\inst{1}
\and R.J. Brunner\inst{1}
\and A.A. Mahabal\inst{1}
\and S.C. Odewahn\inst{2}
\and R.R de Carvalho\inst{3}
\and R.R. Gal\inst{1}
\and P. Stolorz\inst{4}
\and R. Granat\inst{4}
\and D. Curkendall\inst{4}
\and J. Jacob\inst{4}
\and S. Castro\inst{1}
}
\authorrunning{S. G. Djorgovski, \etal}
\institute{
Palomar Observatory, California Inst.~of Technology, Pasadena, CA 91125, USA 
\and 
Dept.~of Physics \& Astronomy, Arizona State Univ., Tempe, AZ 85287, USA
\and 
Observatorio Nacional, CNPq, Rio de Janeiro, Brasil
\and 
Jet Propulsion Laboratory, Pasadena, CA, 91109, USA
}

\maketitle              % typesets the title of the contribution

\begin{abstract}
We review some of the scientific opportunities and technical challenges posed
by the exploration of the large digital sky surveys, in the context of a
Virtual Observatory (VO).  The VO paradigm will profoundly change the way
observational astronomy is done.  Clustering analysis techniques can be used 
to discover samples of rare, unusual, or even previously unknown types of
astronomical objects and phenomena.  Exploration of the previously poorly
probed portions of the observable parameter space are especially promising. 
We illustrate some of the possible types of studies with examples drawn from
DPOSS; much more complex and interesting applications are forthcoming.
Development of the new tools needed for an efficient exploration of these vast
data sets requires a synergy between astronomy and information sciences, with
great potential returns for both fields.
\footnote{
To appear in: Mining the Sky, eds. A. Banday et al., ESO Astrophysics
Symposia, Berlin: Springer Verlag, in press (2001).}
\end{abstract}

\section{Introduction: the Challenges of the Data Abundance}

A paradigm shift is now taking place in astronomy and space science.
Astronomy has suddenly become an immensely data-rich field, with numerous
digital sky surveys across a range of wavelengths, with many Terabytes of
pixels and with billions of detected sources, often with tens of measured
parameters for each object.  We can now map the universe systematically, in
a panchromatic manner.  Even larger, Petabyte-scale astronomical data sets are
now on the horizon. 

This richness of data will enable quantitatively and qualitatively new science,
from statistical studies of our Galaxy and the large-scale structure in the
universe, to the discoveries of rare, unusual, or even completely new types of
astronomical objects and phenomena.  This new data-mining astronomy will also
enable and empower scientists and students anywhere, without an access to large
telescopes, to do first-rate science.  This will invigorate the field, as it
will open the access to unprecedented amounts of data to a fresh pool of
talent. 

In order to cope with this data avalanche, the US astronomical community has
started the National Virtual Observatory (NVO) initiative.  The NVO received
the highest priority recommendation in the National Academy of Sciences decadal
survey, {\em Astronomy and Astrophysics in the New Millennium}. 
This is becoming an international effort, leading to a Global Virtual
Observatory, as it was clearly apparent at this conference and an earlier
meeting in Pasadena (\cf Brunner \etal 2001a). 
\footnote{{\tt http://www.astro.caltech.edu/nvoconf/}}

Full and effective scientific exploitation the vast new data volumes poses 
considerable technical and even deeper, methodological challenges.  The
traditional astronomical data analysis methods are inadequate to cope with this
sudden increase in the {\em data volume} (by several orders of magnitude), and
especially {\em data complexity} (tens or hundreds of dimensions of the
parameter space).  In this review we address the opportunities and problems
posed by the application of automated classification or clustering analysis in
the context of large digital sky surveys, in a future Virtual Observatory (VO).
These challenges require substantive collaborations and partnerships between
astronomy and computer science, promising to bring advances to both fields.

\section{Exploration of the Observable Parameter Space}

A major type of the scientific studies we anticipate for a VO is a systematic
exploration of parameter spaces of measured source attributes from large
digital sky surveys.  This is already done at some level in the catalog domain,
where every source is represented as a point or vector in a multidimensional
parameter space; however, much more ambitious and complex applications are
forthcoming, especially with multiple sky surveys federated within a VO.
In the future, we can also contemplate such explorations in the image or pixel
domain, or in combination of catalog and image domains.  Also, adding the time
axis, from synoptic and sky monitoring surveys, would both literally and
metaphorically add a new dimension for this type of studies.

For example, we can exploit the sheer size of the data sets (billions of
detected sources) to find rare types of objects (\eg one in a million, or one
in a billion, down to some survey flux limit), and use the rich information
content (or complexity) of the data, \ie tens or hundreds of measured
parameters, to achieve optimal discrimination of interesting types of objects
from the more common species.  This includes an exciting possibility of
discovering some previously unknown types of astronomical objects or phenomena.
For a related review, see Djorgovski \etal (2001).

More generally, we see the exploration of observable parameter spaces, created
by combining of large sky surveys over a range of wavelengths, as one of the
chief scientific purposes of a VO.  A complete observable parameter space
axes include quantities such as the object coordinates, velocities or
redshifts, sometimes proper motions, fluxes at a range of wavelength (i.e.,
spectra; imaging in a set of bandpasses can be considered a form of a very
low resolution spectroscopy), surface brightness and image morphological
parameters for resolved sources, variability (or, more broadly, power spectra)
over a range of time scales, \etc~  Any given sky survey samples only a small
portion of this grand observable parameter space, and is subject to its own
selection and measurement limits, \eg limiting fluxes, surface brightness,
angular resolution, spectroscopic resolution, sampling and baseline for
variability if multiple epoch observations are obtained, \etc

Thus, any given sky survey provides only a very limited picture of the
universe, but hopefully with the well understood limitations.  An intelligent
combination of multiple sky surveys within a VO provides a way of overcoming
some of these limitations and enabling a more complete (panchromatic,
multi-scale, synoptic, \etc) view of the physical universe. 

Sometimes simply a combination of sky surveys from very different wavelengths
can produce great new discoveries: recall the discovery of quasars and powerful
radio galaxies as optical IDs of radio sources from the first radio surveys;
or, the discoveries of all kinds of x-ray sources from the optical follow-up of
the early x-ray missions; or the ultraluminous sources found by IRAS; and so
on...  Yet, the multiwavelength studies envisioned for a VO would be far more
comprehensive and ambitious, sampling the previously poorly known portions
of the observable parameter space.

An early vision of such systematic exploration of the observable universe was
promoted by Zwicky (\eg Zwicky 1957), who was -- as usual -- far ahead of his
time, and unfortunately limited by the observational technology available to
him.  Further ideas along these lines have been discussed by Harwit (1975)
and Harwit \& Hildebrand (1986).  These authors recognised that fundamentally
new discoveries can be made by opening of the new portions of the observable
parameter space.  With the plethora of large digital sky surveys now coming
on line, and the technologies developed for their exploration and analysis,
we are now starting to fulfill this vision.

\section{Examples of Science Drivers: Rare and New Types of Objects}

There is already a booming industry of searches for rare, but known types of 
astronomical objects in large digital sky surveys, such as the high-$z$ quasars
or brown dwarfs.  The rarity may be simply the consequence of the observational
selection; for example, brown dwarfs must be very common in the universe, they
are just hard to find.  
With a known type of objects, their properties (\eg typical spectra \etc) can
be convolved with the survey instrumental parameters, such as the bandpasses,
flux limits, \etc, and a particular region of the parameter space where such
objects should be found can be designated for their selection. 
Examples of searches for high-$z$ quasars include, \eg 
Warren \etal (1987),
Irwin \etal (1991),
Kennefick \etal (1995a, 1995b),
Fan \etal (1999, 2000a, 2000c), \etc~ 
The analogous technique is now commonly used to select galaxies at very high
redshifts (\cf Steidel \etal 1999, or Dickinson \etal 2000, and references
therein). 
Examples of searches for brown dwarfs include, \eg
Kirkpatrick \etal (1999),
Strauss \etal (1999),
Burgasser \etal (2000),
Fan \etal (2000b),
Leggett \etal (2000), \etc

In the case of spatially unresolved sources (``starlike'' in optical and NIR),
the only discriminating information between physically different kinds of
sources is in the broad-band spectral energy distribution, which can be
parametrised as a set of the flux ratios (colors) between different bandpasses;
the searches are then done in the color parameter space.  Such photometrically
selected candidates are then followed up spectroscopically. 

This is illustrated in Fig. 1, on an example of color selection of high-$z$ and
type-2 quasars discovered in the Digital Palomar Observatory Sky Survey (DPOSS;
see Djorgovski \etal 1998; and in prep.). 
Normal stars form a temperature sequence, seen here as a banana-shaped locus of
points in this color space.  The spectra of these types of quasars, when folded
through the survey filter curves produce discrepant colors which distinguish
them from those of normal stars. 

\begin{figure}
\centering
\includegraphics[width=.7\textwidth]{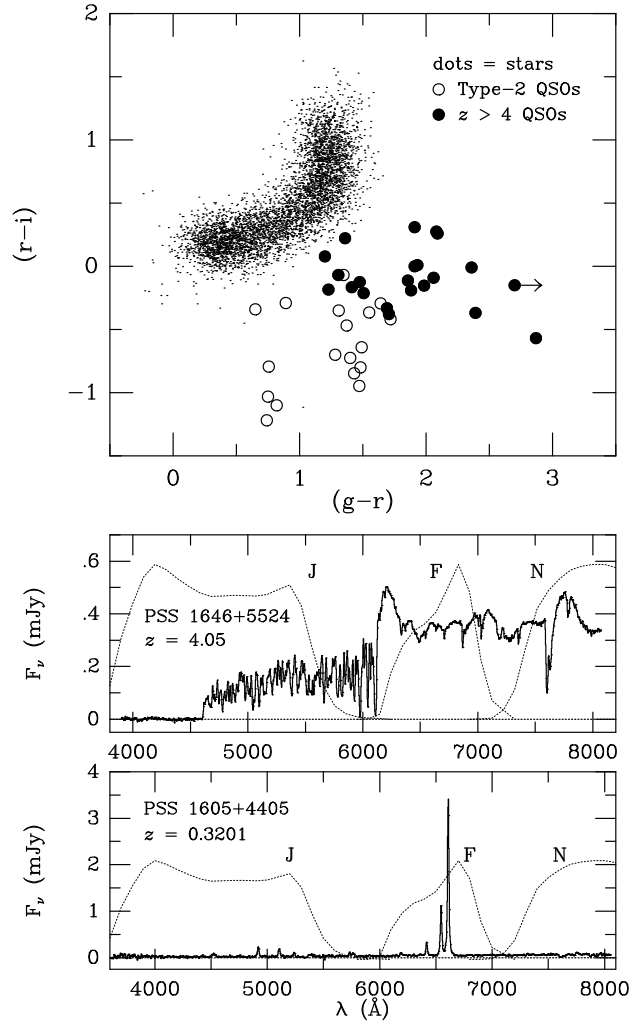} 
\caption[]{
$Top:$~
A representative color-color plot for objects classified as PSF-like in DPOSS.
The dots are normal stars with $r \sim 19$ mag.  Solid circles are some of the
$z > 4$ quasars, and open circles are some of the type-2 quasars found in this
survey. 
~$Middle:$~
A spectrum of a typical $z > 4$ quasar, with the DPOSS bandpasses shown as the
dotted lines.  The mean flux drop blueward of the Ly$\alpha$ line, caused by
the absorption by Ly$\alpha$ forest and sometimes a Lyman-limit system, gives
these objects a very red $(g-r)$ color, while their intrinsic blue color is
retained in $(r-i)$, placing them in the lower right portion of this color-color
diagram.
~$Bottom:$~
A spectrum of a typical type-2 quasar, with the DPOSS bandpasses shown as the
dotted lines.  The presence of the strong [O III] lines is the $r$ band places
such objects below the stellar locus in this color-color diagram.
\label{fig1}}
\end{figure}

In the case of high-$z$ quasars, absorption by the intergalactic hydrogen
clouds and gas-rich (proto)galaxies produces a strong drop blueward of the
quasar's own Ly$\alpha$ emission line center, and thus a very red $(g-r)$
color, while the observed $(r-i)$ color reflects the intrinsically blue
spectrum of the quasars: these objects are ``red in the blue, and blue in the
red'', unlike any ordinary stars.  To date, $\sim 100$ such quasars have been
found in DPOSS. 
\footnote{
We make them publicly available through our webpage,\\
{\tt http://www.astro.caltech.edu/$\sim$george/z4.qsos}}

In the case of type-2 quasars the with hidden central engines (nonthermal
continuum and broad-line regions) but with the (mostly?) unobscured narrow-line
regions, presence of the strong, narrow emission lines in one of the survey
bands can produce peculiar colors.  For the type-2 quasars discovered in DPOSS
(Djorgovski \etal 1999, and in prep.), the strong [O III] lines traverse the
$r$ band in the redshift interval $z \sim 0.31 - 0.38$, and separate these
objects in the color space away from the stellar locus.

Both of these types of objects are relatively rare, with surface densities
$\lsim 10^{-2}$ deg$^{-2}$ down to the reliable star-galaxy classification
limit in DPOSS, i.e., down to $r \sim 19.5$ mag.  Thus, one must have a survey
covering both a large area, and going sufficiently deep, in order to detect
statistically meaningful samples of them, as well as the suitable selection
methodology.

A similar approach is used in surveys with a poor angular resolution at other
wavelengths, e.g., as a star-galaxy separation in IRAS data (Boller \etal 1992),
or as a way of distinguishing likely quasars from radio galaxies using a
spectral index in the radio, or using the x-ray hardness ratio to separate AGN
from other types of sources, \etc 
Complete samples of quasars at all redshifts can be generated using similar
approaches in the optical and NIR (\cf Wolf \etal 1999, and this volume; or
Warren \etal 2000).  Likewise, stars of a particular spectral type can be
selected and used as probes of the Galactic structure (\eg Yanny \etal 2000). 
If morphological parameters of resolved galaxies can be parametrised in some
suitable manner, the same approach can be used to isolate galaxies in some
range of Hubble types (\eg Odewahn \etal 1996). 

However, an even more intriguing prospect is the discovery of rare and
previously $unknown$ types of objects in these large data sets, which may have
been missed so far due to the rarity, and/or the blending with some more
familiar types (\eg all unresolved sources look alike on images).  They may be
uncovered through a systematic search for outliers in some parameter space, \ie
as objects empirically distinct from ``everything else'' in a statistically
quantifiable manner.  

Possible examples of new kinds of objects (or at least extremely rare or
peculiar sub-species of known types of objects) have been found in the course
of high-$z$ quasar searches by both SDSS (Fan \& Strauss, priv. comm.) and
DPOSS groups.  Some examples from DPOSS are shown in Figure 2.  These objects
have most unusual, and as yet not fully (or not at all) understood spectra,
which cause them to have peculiar broad-band colors, which serendipitously
place them in the region of the color space where high-$z$ quasars are to be
found.  

\begin{figure}
\centering
\includegraphics[width=.9\textwidth]{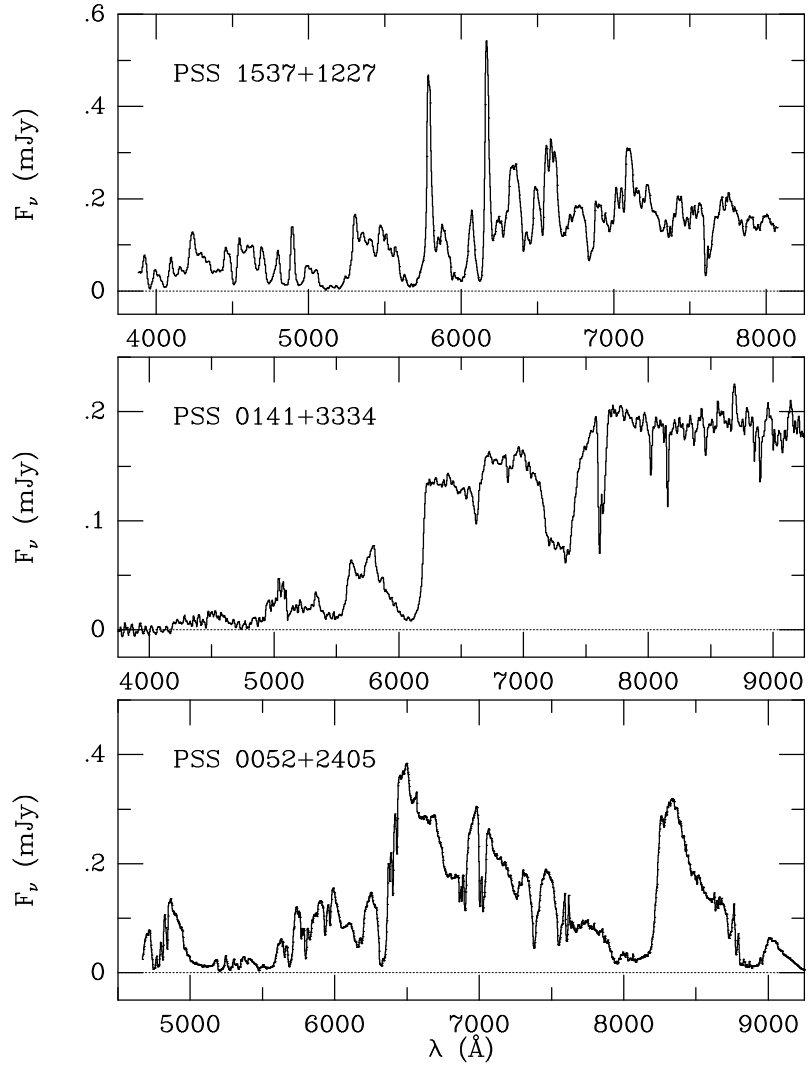} 
\caption[]{
Spectra of some of the peculiar objects discovered in DPOSS during the high-$z$
quasar search.
The top one, PSS 1537+1227, has been identified as an extreme case of a rare
type of a low-ionisation, Fe-rich, BAL QSO, at $z \approx 1.2$.  A prototype
case (but with a spectrum not quite as extreme as this) is FIRST 0840+3633,
discovered by Becker \etal (1997).
The nature of the other two remains uncertain as of this writing, but it is
possible that they are also some peculiar sub-species of BAL QSOs;
but perhaps they are something else entirely.
\label{fig2}}
\end{figure}

A systematic search for outliers in other, as yet unexplored portions of this
parameter space may yield additional peculiar objects, some of which may turn
out to be prototypes of new astrophysical phenomena.  A thorough, large-scale,
unbiased, multi-wavelength census of the universe will any such new types of
objects and phenomena, if they do exist and are detectable in the available
data.  These may be exciting new discoveries with a VO. 

In addition to the searches for the rare, natural phenomena, this methodology
(and perhaps also some of the VO data sets) can form a basis for a generalised
and more powerful approach to SETI (Djorgovski 2000).

\section{Clustering Analysis Challenges in a Virtual Observatory}

Separation of the data into different types of objects, be it known or unknown
in nature, can be approached as a problem in automated classification or
clustering analysis.  This is a part of a more general and rapidly growing
field of Data Mining (DM) and Knowledge Discovery in Databases (KDD).  We see
here great opportunities for collaborations between astronomers and computer
scientists and statisticians.  For an overview of some of the issues and
methods, see the volume edited by Fayyad \etal (1996b), as well as several
papers in this volume.

If applied in the catalog domain, the data can be viewed as a set of $n$ points
or vectors in an $m$-dimensional parameter space, where $n$ can be in the range
of many millions or even billions, and $m$ in the range of a few tens to
hundreds.  The data may be clustered in $k$ statistically distinct classes,
which could be modeled, \eg as multivariate Gaussian clouds, and which
hopefully correspond to physically distinct classes of objects (\eg stars,
galaxies, quasars, \etc).  This is schematically illustrated in Figure 4.

\begin{figure}
\centering
\includegraphics[width=.7\textwidth]{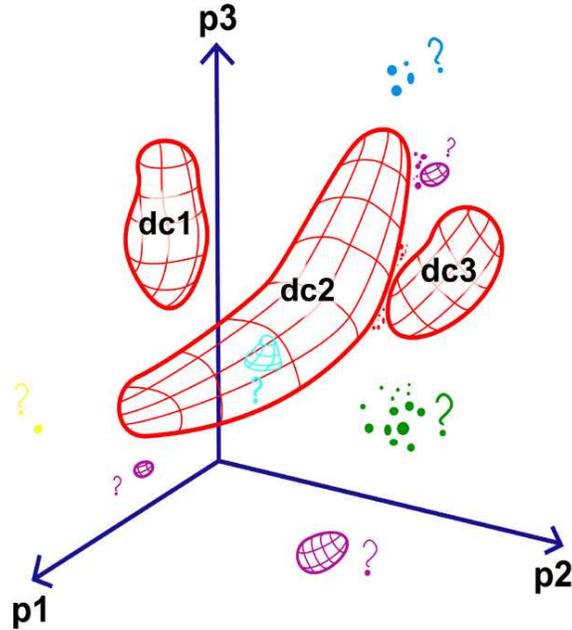} 
\caption[]{
A schematic illustration of the problem of clustering analysis in some
parameter space.  In this example, there are 3 dimensions, $p1$, $p2$, and $p3$
(\eg some flux ratios or morphological paremeters), and most of the data points
belong to 3 major clusters, denoted $dc1$, $dc2$, and $dc3$ (\eg stars,
galaxies, and ordinary quasars).  One approach is to isolate these major
classes of objects for some statistical studies, \eg stars as probes of the
Galactic structure, or galaxies as probes of the large scale structure of the
universe, and filter out the ``anomalous'' objects.  A complementary view is to
look for other, less populated, but statistically significant, distinct
clusters of data points, or even individual outliers, as possible examples of
rare or unknown types of objects.  Another possibility is to look for holes
(negative clusters) within the major clusters, as they may point to some
interesting physical phenomenon -- or to a problem with the data.
\label{fig3}}
\end{figure}

A typical VO data set may have the following properties: $\sim 10^9$ data
vectors in $\sim 10^2$ dimensions (these are measured source attributes,
including positions, fluxes in different bandpasses, morphology quantified
through different moments of light distribution and other suitably constructed 
parameters, \etc).  Some of the parameters would be primary measurements, and
others may be derived attributes, such as the star-galaxy classification (e.g.,
from a supervised classifier such as an Artificial Neural Net, or some Bayesian
scheme), some may be ``flags'' rather than numbers, some would have error-bars
associated with them, and some would not, and the error-bars may be functions
of some of the parameters, e.g., fluxes. Some measurements would be present
only as upper or lower limits. Some would be affected by ``glitches'' due to
instrumental problems, and if a data set consists of a merger of two or more
surveys, e.g., cross-matched optical, infrared, and radio (and this would be a
common scenario within a VO), then some sources would be misidentified, and
thus represent erroneous combinations of subsets of data dimensions.  Surveys
would be also affected by selection effects operating explicitly on some
parameters (e.g., coordinate ranges, flux limits, \etc), but also mapping onto
some other data dimensions through correlations of these properties; some
selection effects may be unknown. 

Physically, the data set may consist of a number of distinct classes of
objects, such as stars (including a range of spectral types), galaxies
(including a range of Hubble types or morphologies), quasars, \etc  Within each
object class or subclass, some of the physical properties may be correlated,
and some of these correlations may be already known and some as yet unknown,
and their discovery would be an important scientific result by itself.
Correlations of independently measured physical parameters represent a
reduction of the statistical dimensionality in a multidimensional data
parameter space, and their discovery may be an integral part of the clustering
analysis.

If the number of object classes $k$ is known (or declared) {\it a priori}, and
training data set of representative objects is available, the problem reduces
to supervised classification, where tools such as Artificial Neural Nets or
Decision trees can be used.  This is now commonly done for star-galaxy
separation in the optical or NIR sky surveys (\eg Odewahn \etal 1992, or
Weir \etal 1995).  Searches for known types of objects with predictable
signatures in the parameter space (\eg high-$z$ quasars) can be also cast in
this way. 

However, a more interesting and less biased approach is where the number of
classes $k$ is not known, and it has to be derived from the data themselves.
The problem of unsupervised classification is to determine this number in some
objective and statistically sound manner, and then to associate class
membership probabilities for all objects.  Majority of objects may fall into a
small number of classes, \eg normal stars or galaxies.  What is of special
interest are objects which belong to much less populated clusters, or even
individual outliers with low membership probabilities for any major class.
Some initial experiments with unsupervised clustering algorithms in the
astronomical context include, \eg Goebel \etal (1989), Weir \etal (1995), de
Carvalho \etal (1995), and Yoo \etal (1996), but a full-scale application to
major digital sky surveys yet remains to be done.  
Intriguing applications which addressed the issue of how many statistically
distinct classes of GRBs are there include Mukherjee \etal (1998) and
Rogier \etal (2000).
One method we have been experimenting with (applied on the various data sets
derived from DPOSS) is the Expectation Maximisation (EM) technique, with the
Monte Carlo Cross Validation (MCCV) as the way of determining the maximum
likelihood number of the clusters. 
An array of good unsupervised classification techniques will be an essential
part of a VO toolkit.  

This may be a computationally very expensive problem. 
For the simple $K$-means algorithm, the computing cost scales as
$K ~\times ~N ~\times ~I ~\times ~D$,
where
$K$ is the number of clusters chosen {\it a priori},
$N$ is the number of data vectors (detected objects),
$I$ is the number of iterations,
and
$D$ is the number of data dimensions (measured parameters per object).
For the more powerful Expectation Maximisation technique, the cost scales as
$K ~\times ~N ~\times ~I ~\times ~D^2$, 
and again one must decide {\it a priori} on the value of $K$.  If this number
has to be determined intrinsically from the data, \eg with the Monte Carlo
Cross Validation method, the cost scales as
$M ~\times ~K_{max}^2 ~\times ~N ~\times ~I ~\times ~D^2$
where
$M$ is the number of Monte Carlo trials/partitions,
and
$K_{max}$ is the maximum number of clusters tried.
Even with the typical numbers for the existing large digital sky surveys
($N \sim 10^8 - 10^9$, $D \sim 10 - 100$) this is already reaching in the
realm of Terascale computing, especially in the context of an interactive and
iterative application of these analysis tools.  Development of faster and
smarter algorithms is clearly a priority.

One technique which can simplify the problem is the multi-resolution
clustering.  In this regime, expensive parameters to estimate, such as the
number of classes and the initial broad clustering are quickly estimated using
traditional techniques, and then one could proceed to refine the model locally
and globally by iterating until some objective statistical (\eg Bayesian)
criterion is satisfied. 

One can also use intelligent sampling methods where one forms ``prototypes''of
the case vectors and thus reduces the number of cases to process.  Prototypes
can be determined from simple algorithms to get a rough estimate, and then
refined using more sophisticated techniques.  A clustering algorithm can
operate in prototype space.  The clusters found can later refined by locally
replacing each prototype by its constituent population and reanalyzing the
cluster. 

Techniques for dimensionality reduction, including principal component analysis
and others can be used as preprocessing techniques to automatically derive the
dimensions that contain most of the relevant information. 

There are many other technical and methodological challenges in this quest,
primarily the problems steming from the heterogeneity and intrinsic complexity 
of the data, including treatment of upper an lower limits, missing data,
selection effects and data censoring, \etc  These issues affect the proper
statistical description of the data, which then must be reflected in the
clustering algorithms.

Related to this are the problems arising from the data modeling.  The commonly
used assumption of clusters represented as multivariate Gaussian clouds is
rarely a good descriptor of the reality.  Clusters may have non-Gaussian
shapes, \eg exponential or power-law tails, asymmetries, sharp cutoffs, \etc~
This becomes a critical issue in evaluating the membership probabilities in
partly overlapping clusters, or in a search for outliers (anomalous events) in
the tails of the distributions.  In general, the proper functional forms for
the modeling of clusters are not known {\it a priori}, and must be discovered
from the data.  Applications of non-parametric techniques may be essential
here. A related, very interesting problem is posed by the {\it topology} of
clustering, with a possibility of multiply-connected clusters or gaps in the
data (\ie negative clusters embedded within the positive ones), hierarchical or
multi-scale clustering (\ie clusters embedded within the clusters) \etc 

The clusters may be well separated in some of the dimensions, but not in others.
How can we objectively decide which dimensions are irrelevant, and which ones
are useful?  An automated and objective rejection of the ``useless'' dimensions,
perhaps through some statistically defined enthropy criterion, could greatly
simplify and speed up the clustering analysis. 

Once the data are partitioned into distinct clusters, their analysis and
interpretation starts.  One question is, are there interesting (in general,
multivariate) correlations among the properties of objects in any given
cluster?  Such correlations may reflect interesting new astrophysics (\eg, the
stellar main sequence, the Tully-Fisher and Fundamental Plane correlations for
galaxies, \etc), but at the same time complicate the statistical interpretation
of the clustering.  They would be in general restricted to a subset of the
dimensions, and not present in the others.  How do we identify all of the
interesting correlations, and discriminate against the ``uninteresting''
observables? 

Given these issues, a blind applications of the commonly used (commercial or
home-brewed) clustering algorithms in such real-life cases could produce some
seriously misleading or simply wrong results.  The clustering methodology must
be robust enough to cope with these problems, and the outcome of the analysis
must have a solid statistical foundation. 

Effective and powerful data visualization, applied in the parameter space
itself, is another essential part of the interactive clustering analysis.  
Good visualisation tools are also essential for the interpretation of results,
especially in an iterative environment.  While clustering algorithms can assist
in the partitioning of the data space, and can draw the attention to anomalous
objects, ultimately a scientist guides the experiment and draws the conclusions.

Another key issue is interoperability and reusability of algorithms and models
in a wide variety of problems posed by a rich data environment such as
federated digital sky surveys in a VO. 

Finally, a scientific verification and evaluation, testing, and follow-up on
any of the newly discovered classes of objects, physical clusters discovered by
these methods, and other astrophysical analysis of the results is essential in
order to demonstrate the actual usefulness of these techniques for a VO or
other applications.  Clustering analysis can be seen as a prelude to the more
traditional type of astronomical studies, as a way of selecting of interesting
objects of samples.

\section{Exploration of the Time Domain and the Image Domain}

The current generation of digital sky surveys provides ``extended snapshots''
of the sky, at some set of wavelengths.  But the future brings the ``movies''.
A major new area for exploration will be in the time domain, with a number of
ongoing or forthcoming surveys aiming to map large portions of the sky in a
repeated fashion, down to very faint flux levels. 
In addition to a large number of ongoing microlensing surveys, searches for
Solar system objects, supernova surveys, and other searches for variable or
transient sources at optical wavelengths, many more are being planned or
proposed; the ultimate such experiment would be the Large Synoptic Survey
Telescope, which received a high recommendation in the NAS decadal survey. 
The subject is reviewed, \eg by Paczy\'nski (2000; and this volume) and 
Diercks (2001).

\begin{figure}
\centering
\includegraphics[width=.4\textwidth]{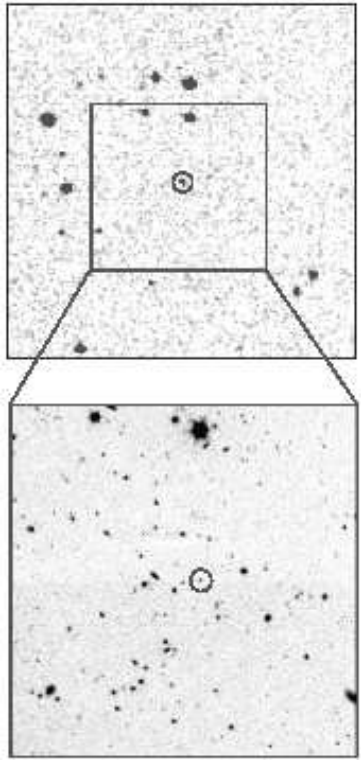} 
\caption[]{
A serendipitously discovered optical transient event PVO 1213+0903
(PVO is a designation for a Palomar Variable Object, selected from DPOSS).
Top:
A portion of a DPOSS $F$ plate image with an $r \sim 18.5$ mag, starlike object,
circled.  The object was selected due to its apparent peculiar color (bright in
$r$, extremely faint in the other two DPOSS bands); however, this was simply a
consequence of the plates taken at different times, with one of them catching
it in a bright state. 
Bottom:
A portion of the corresponding Keck $R$ band image.  The DPOSS transient was
positionally coincident with an $R \sim 24.5$ mag galaxy, with an estimated
probable $z \sim 1$.  At such a redshift, this object would have been a few
hundred times brighter than a supernova at its peak.  It may be an example of a
GRB ``orphan afterglow'', or possibly some other, new type of a transient. 
\label{fig4}}
\end{figure}

\begin{figure}
\centering
\includegraphics[width=.8\textwidth]{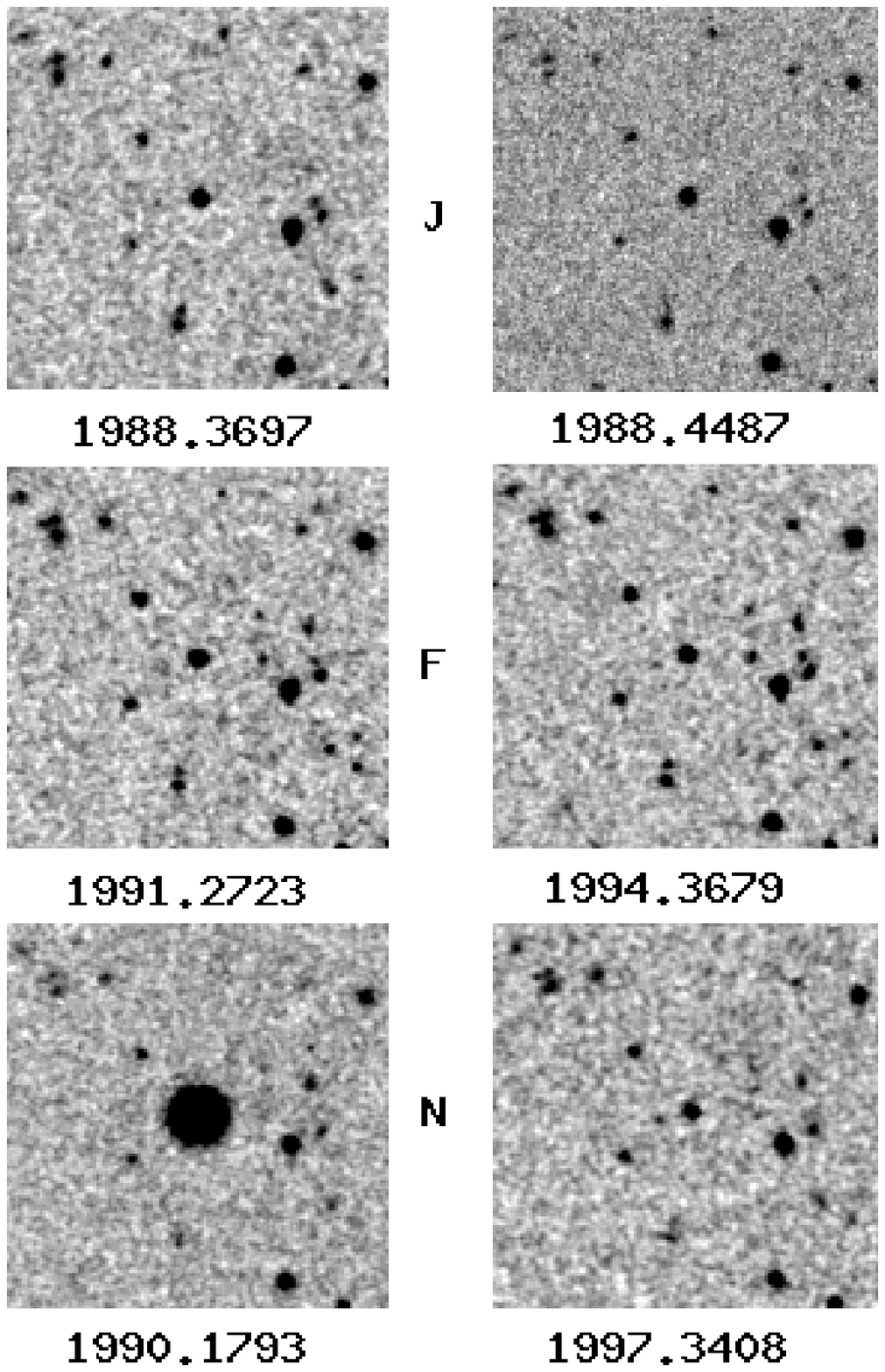} 
\caption[]{
A star which went BANG! in the night...
Images of a star, PVO 1558+3725, seen in the plate overlaps in $J$ (= green,
top) $F$ (= red, middle) and $N$ ($\approx i$ band, bottom).  The observation
epochs are indicated below each panel.  The star was brighter by at least a
factor of 300 on the $N$ plate taken on 1990.1793 UT.  Its subsequent
spectroscopy shows normal, early-type absorption spectrum, with no line
emission.  The cause, amplitude, and duration of the outburst are unknown. How
many normal stars do this, how often, and why?  New synoptic sky surveys may
help answer this. 
\label{fig5}}
\end{figure}

While these surveys may start generating Petabytes of data, they will open 
a whole new field of searches for variable astronomical objects.  By analogy
with the searches for rare types of objects in the image domain, we can expect
both assembly of large, statistical samples of known types of variables
(\eg variable stars of all kinds, supernov\ae, and AGN), as well as possible
new types of variable or transient objects.  We know surprisingly little about
the faint, variable sky at any wavelength, and over most time scales.  A 
panchromatic approach to the variable universe is long overdue.  DM techniques 
described above can be applied directly to the analysis of such data. 

As an illustration of the kind of unexpected phenomena which may be found, we
show here a couple of examples drawn from DPOSS.  The POSS-II plates in
different filters are taken at different times, so that highly variable objects
would appear as having peculiar colors (\eg much brighter in one band than in
the others).  Also, about 50\% of the northern sky which DPOSS covers is imaged
at least twice in each band due to plate overlaps.  Figure 4 shows an example
of a serendipitously discovered optical transient associated with a faint
galaxy.  Figure 5 shows a star seen in a plate overlap region (\ie photographed
twice in each of the 3 bands), which was brighter by at least a factor of 300
during one of the exposures (it is a lower limit, since the image is saturated,
and also the outburst may have lasted less than the plate exposure time, but it
is averaged over it). These extreme events have been picked up serendipitously
in the course of other DPOSS work.  A systematic search for variable or
transient sources is likely to yield many more. 

\begin{figure}
\centering
\includegraphics[width=.9\textwidth]{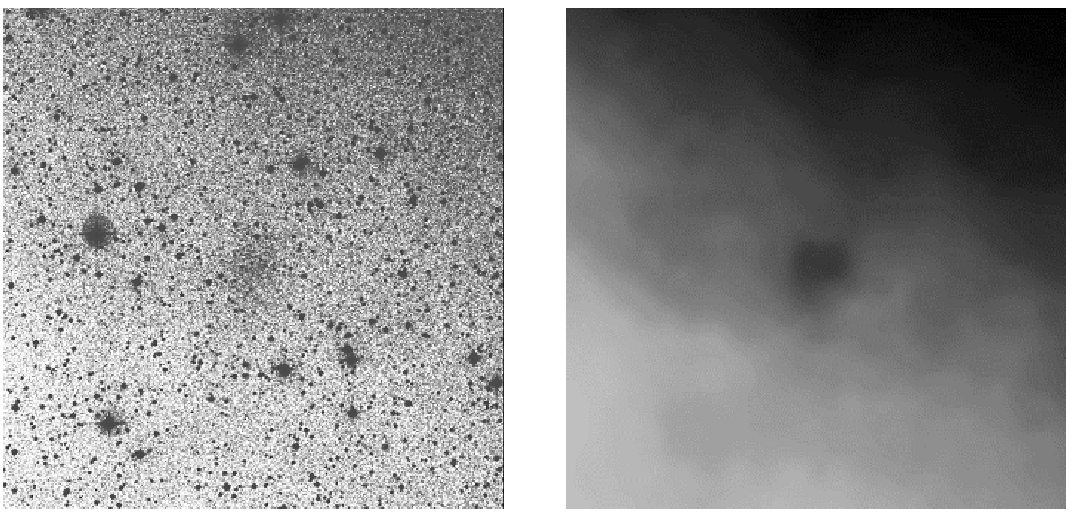} 
\caption[]{
A demonstration of a detection process for detection of LSB sources missed by
most survey pipelines, from Brunner \etal (2001b).  On the left is the original
DPOSS plate image which contains the dwarf spheroidal galaxy Andromeda V
(Armandroff \etal 1998), which is visible with a large image stretch.  On the
right is a filtered version of the original image which is designed to
emphasize subtle background variations which can be caused by low surface
brightness sources. 
\label{fig6}}
\end{figure}

Likewise, the low surface brightness universe (at any wavelength!) remains as
one of the remaining frontiers.  For reviews, see, \eg Impey \& Bothun (1997),
or Schombert (2001), and references therein.  In a more general sense, we can
cast the problem as the detection of sources at a range of surface brightness
contrasts over a range of scales, and possibly with a range of image
morphologies.  Practically every implemented source detection algorithm has
a preferred range of angular scales and a limiting surface brightness at any
given angular scale.  Exploration of novel source or structure detection
algorithms seems to be in order.  Broadening of the dynamical range of the
limiting surface brightness and software-limited angular resolution in digital
sky surveys at any wavelength would be an important area of work in a VO.

As a demonstration of the type of science which is facilitated by a virtual
observatory, we have undertaken a project utilizing both images and catalogs to
explore the multi-wavelength, low surface brightness universe (Brunner \etal
2001b; and in prep.; see also Testa \etal, this volume).  Our analysis
techniques are complimentary to normal data reduction pipeline techniques in
that we focus on the diffuse emission that is ignored or removed by more
traditional algorithms.  This requires a spatial filtering which must account
for objects of interest, in addition to observational artifacts (\eg bright
stellar halos).  With this work we are exploring the intersection of the
catalog and image domains in order to maximize the scientific information we
can extract from the federation of large survey data. 

Additional field to explore may be the use of automated pattern recognition
algorithms applied in the image domain, rather than in some parameter space of
derived object attributes, \ie the catalog domain (thus perhaps bypassing the
tricky problem of source detection).  
Such ``artificial vision'' tools may be used to discover sources with a
particular image morphology (\eg galaxies of a certain type). An example from
planetary science, an automated discovery of volcanos in Magellan Venus radar
images, was described in Fayyad \etal (1996a) and Burl \etal (1998).  An even
more interesting approach would be to employ AI techniques to search through
panoramic images (perhaps matched from multiple wavelengths) for unusual image
patterns, possibly correlated with some data context (\eg always found in or
near a cluster of galaxies, or a molecular cloud, or coincident with a spiral
galaxy, \etc).  For example, it may be possible for a program to find features
such as the gravitationally lensed arcs in rich clusters (perhaps in a way
which may mimic the discovery process in the minds of Lynds \& Petrosian 1989),
but possibly also some other, as yet unknown phenomena. 

\section{Concluding Comments: Towards a Virtual Observatory}

The technical problems posed by the analysis of the VO-related data sets are
considerable, but within our reach.  These problems are common to all
data-intensive fields today.  However, we believe that astronomy is ``just
right'' as a testbed for these computing methodologies: the size and complexity
of the data sets is nontrivial, but is manageable, providing rewarding
challenges for applied information science.  We thus envision a continuous,
powerful synergy of astronomy and information sciences in tackling these
challenges. 

The scientific applications will not be lacking: once the means are available,
both the data and the exploration tools, some exciting science will result,
including many things we have never thought about.  One of the most important
aspects of a VO is the opening of the field to a broader, world-wide pool of
talent, including many scientists and students without a ready access to the
front-line observational facilities, but with good ideas.

The ultimate, long-term future of observational astronomy may be in pushing the
observations of the universe along all of the physically accessible axes of the
parameter space, over the entire available electromagnetic spectrum and other
information channels, \eg the neutrinos, gravity waves, cosmic rays, clever
uses of gravitational lensing to ``observe'' the dark matter, \etc~ 
In principle we could be doing it down to the physical limits such as the
quantum noise or opacity of the Galactic ISM -- and in the form of a continuous
monitoring of the entire sky, at all wavelengths.  
Of course, there will always be some technical limits (\eg angular diffraction
resolution limits due to the physical size of available telescopes or arrays,
detector technology, \etc), and practical limits (\eg cost). 
This may sound like an insanely ambitious vision today, but think what would,
say, any good, early 20th century astronomer think about the array of
telescopes, sky surveys, and various tools at our disposal today.  If there is
any lesson in the past, then it is that we cannot even imagine what will be
possible to do, what will be found, and what will be interesting to do a few
decades from now. 

Astronomy has always pushed the limits of technology for its purposes, be it
optics, telescope design, or detectors.  We are now doing the same with the
information technology.  The VO concept opens a path towards a systematic,
complete exploration of the universe, and gives us a preview and a vision of
the astronomy of the future.

\bigskip
\noindent{\bf Acknowledgements.}~
Prototyping VO developments at Caltech and JPL have been funded by grants from
NASA, the Caltech President's Fund, and several private donors. 
The processing and initial exploration of DPOSS was supported by a generous
gift from the Norris foundation, and by other private donors.  
We are grateful to all people who helped with the creation of DPOSS and with
our Palomar and Keck observing runs, and especially a number of excellent
Caltech undergraduates who worked with us through the years. 
Finally, we wish to acknowledge stimulating interactions with other VO
enthusiasts from Pasadena and other corners of the planet.

%INDEX%%%%%%%%%%%%%%%%%%%%%%%%%%%%%%%%%%%%%%%%%%%%%%%%%%%%%%%%%%%%%%%
\clearpage
\addcontentsline{toc}{section}{Index}
\flushbottom
\printindex
%%%%%%%%%%%%%%%%%%%%%%%%%%%%%%%%%%%%%%%%%%%%%%%%%%%%%%%%%%%%%%%%%%%%%

\end{document}